\documentstyle[aps,preprint]{revtex}

\begin{document}
\draft
\preprint{ISSP Sep. 20, 1994}
\title{Universal Behavior of Correlations between Eigenvalues of Random
Matrices}
\author{T.~S.~Kobayakawa${}^*$, Y.~Hatsugai${}^{**}$, M.~Kohmoto and
  A.~Zee${}^{***}$}
\address{
Institute for Solid State Physics,   University of Tokyo \\
  7-22-1, Roppongi, Minato-ku, Tokyo 106, JAPAN \\
  ${}^{***}$ Institute for Theoretical Physics,   University of California \\
  Santa Barbara, CA 93106, USA
}

\maketitle
\begin{abstract}
The universal connected correlations proposed recently
between eigenvalues of unitary random matrices is examined numerically.
We perform an ensemble average by the Monte Carlo sampling.
Although density of eigenvalues and a bare correlation of the eigenvalues are
not universal, the connected correlation shows a universal behavior
after smoothing.
\end{abstract}
\pacs{}
\narrowtext

Br\'ezin and Zee\cite{pa.bz1,pa.bz2,pa.bz3} have discovered recently
the universal behavior of the eigenvalue correlation of random matrices.
Let $\phi$ be $N \times N$ hermitian matrices
and let $i$th eigenvalue of $\phi$ be denoted by $\lambda_i, \mu_i$ or
$\nu_i$.
We consider the probability distribution with a weight
\begin{equation}
P(\phi) \propto \exp(-N\mathop{\rm{Tr}} V(\phi))
  \label{la.dist}
\end{equation}
where $V(\phi)$ is an even polynomial of $\phi$.
The density of the eigenvalues $\rho(\lambda)$
and the correlation of the eigenvalues $\rho(\mu,\nu)$ are defined as
\begin{eqnarray}
  \rho(\lambda) &=&\langle{1 \over N} \sum_{i=1}^N \delta(\lambda-
\lambda_i)\rangle\\
  \label{la.defdist}
  \rho(\mu,\nu) &=& \langle{1 \over N} \sum_{i=1}^N \delta(\mu-\mu_i)
                   {1 \over N} \sum_{j=1}^N \delta(\nu-\nu_i)\rangle
  \label{la.defcorr}
\end{eqnarray}
where $\langle \ldots \rangle$ is an ensemble average.
Quantities $\rho(\lambda)$ and $\rho(\mu,\nu)$ are not universal
and depend on the details of $P(\phi)$, {\it i.e.} $V(\phi)$.\cite{pa.bipz}
Using orthogonal polynomials, one can calculate
the correlation of the eigenvalues of random matrices.
Using an ansatz for a form of the orthogonal polynomials,
Br\'ezin and Zee\cite{pa.bz1} have discussed a connected part of
$\rho(\mu,\nu)$, which is defined by
\begin{equation}
  \rho_c(\mu,\nu) = \rho(\mu,\nu) - \rho(\mu) \rho(\nu)
  \label{la.defccorr}
\end{equation}
This $\rho_c(\mu,\nu)$ oscillates wildly as a function of $\mu$ and $\nu$.
They have shown that, in the large $N$ limit, the smoothed connected
correlation $\rho_c^{\rm smooth}(\mu,\nu)$ has an universal form
\begin{equation}
  \rho_c^{\rm smooth}(\mu,\nu) = -\frac{1}{2 N^2 \pi^2 a^2}f(\mu/a,\nu/a),
  \label{la.univ}
\end{equation}
with
\begin{equation}
  f(x,y) = \frac{1}{(x-y)^2} \frac{(1-xy)}{\sqrt{(1-x^2)(1-y^2)}}
  \label{la.f}
\end{equation}
where $\pm a$ is the endpoint of the spectra of the eigenvalues.
The smoothed connected correlation $\rho_c^{\rm smooth}(\mu,\nu)$ must be
averaged over intervals $\delta\mu$ and $\delta\nu$ much less than $O(1)$
but larger than $O(1/N)$.
Thus (\ref{la.univ}) is valid where $|\mu-\nu|, |\pm a-\mu|$
and $|\pm a-\nu|$ are larger than $O(1/N)$.
The smoothed connected correlation $\rho_c^{\rm smooth}(\mu,\nu)$ is
universal in the sense that the function $f(x,y)$ does not depend on
the potential $V$ at all.
This universality has since been verified and derived by Beenakker \cite{bee}
and by Eynard\cite{eyn} using other methods.

In this paper we study numerically the correlation of the eigenvalues to
check the
validity of the universality in equations (\ref{la.univ}) and (\ref{la.f}).
In our calculations, we limit the form of $V(\phi)$ to
\begin{equation}
  V(\phi) = v_2 \phi^2 + v_4 \phi^4 + v_6 \phi^6
  \label{la.V}
\end{equation}
We choose the matrix size $N$ to be 100 and performed the ensemble
average
by Monte Carlo~(MC) method with importance sampling.
In the calculation, $10^6$ MC samples are taken.

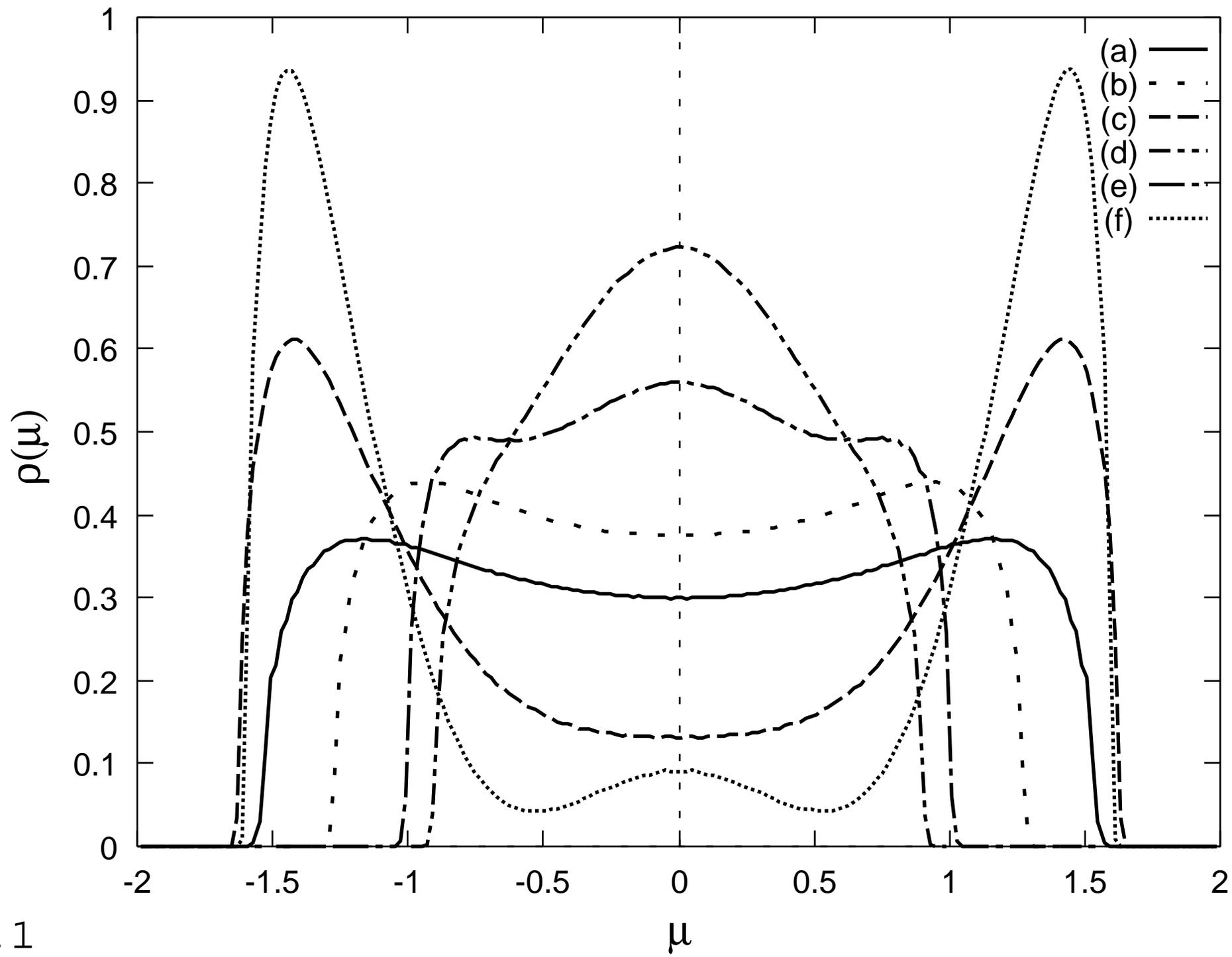
\begin{figure}
  %\epsfile{file=dens.eps,width=8.5cm}
  \caption{The density of the eigenvalues for the probability
    distribution~(1).
    The parameters $(v_2,~v_4,~v_6)$ are :(a)~(0.46,~0.25,~0.21),
    (b)~(0.87,~0.36,~0.68), (c)~(-0.70,~-0.89,~0.91),
    (d)~(5.77,~-5.37,~5.62), (e)~(3.32,~-3.72,~5.20)
    and (f)~(0.0271,~-4.63,~2.55).}
  \label{fig:dens}
\end{figure}

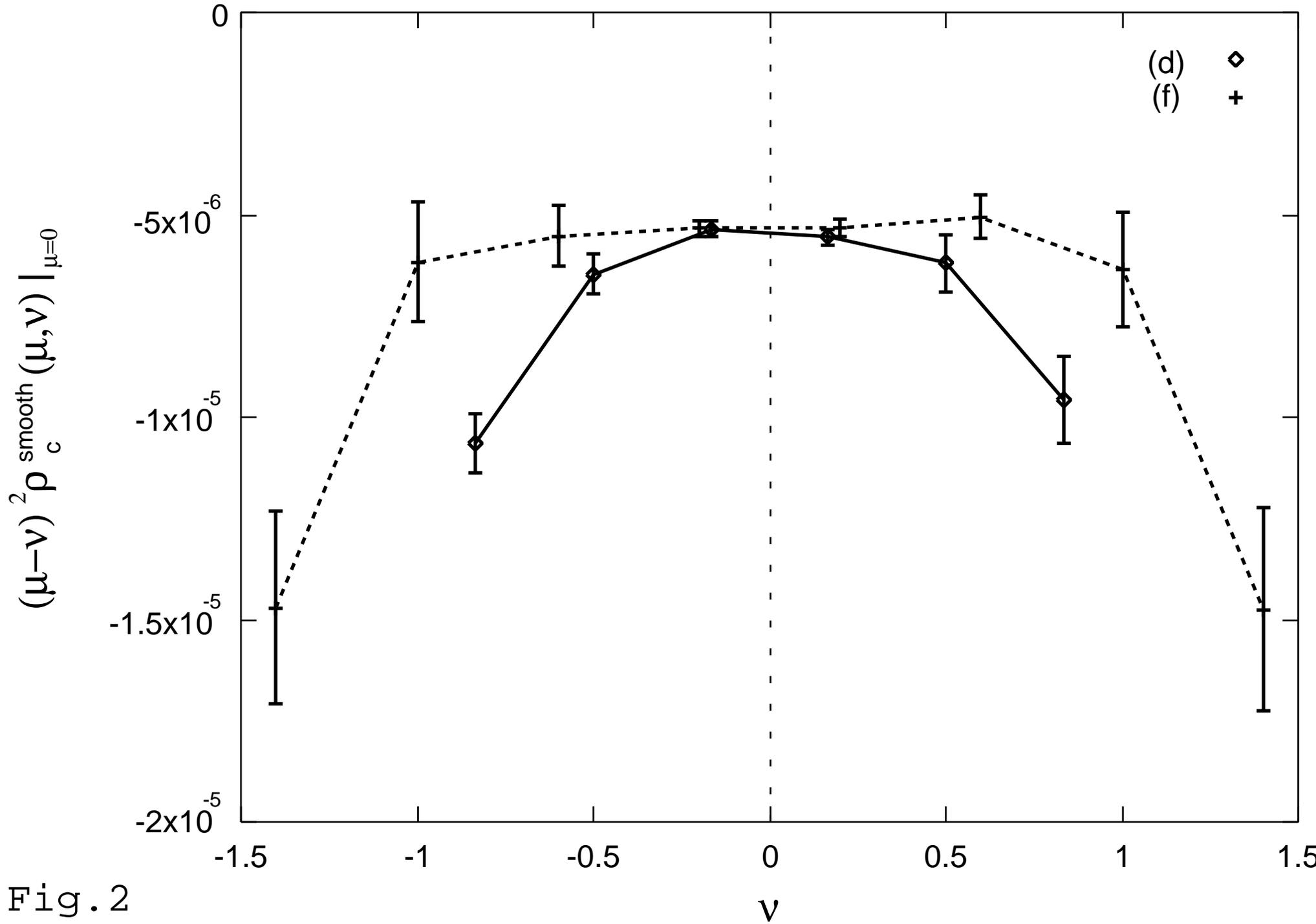
\begin{figure}
  %\epsfile{file=c1.eps,width=8.5cm}
  \caption{The smoothed connected correlation
    $(\mu-\nu)^2\rho_c^{\rm smooth}(\mu,\nu)|_{\mu=0}$.
    The parameters $(v_2,~v_4,~v_6)$ are the same as in Fig.~1.}
  \label{fig:c1}
\end{figure}

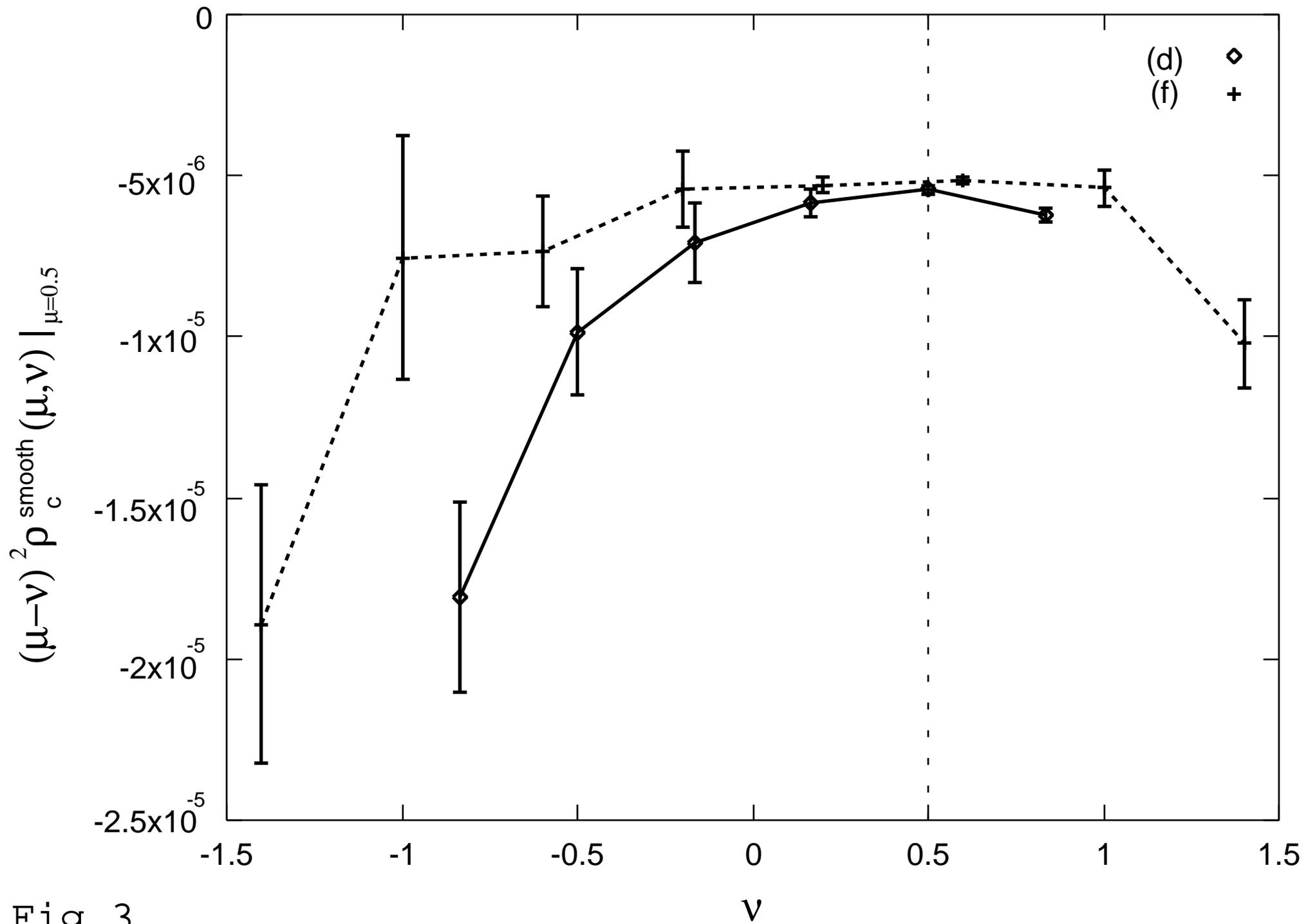
\begin{figure}
  %\epsfile{file=c2.eps,width=8.5cm}
  \caption{The smoothed connected correlation
    $(\mu-\nu)^2\rho_c^{\rm smooth}(\mu,\nu)|_{\mu=0.5}$.
    The parameters $(v_2,~v_4,~v_6)$ are the same as in Fig.~1.}
  \label{fig:c2}
\end{figure}

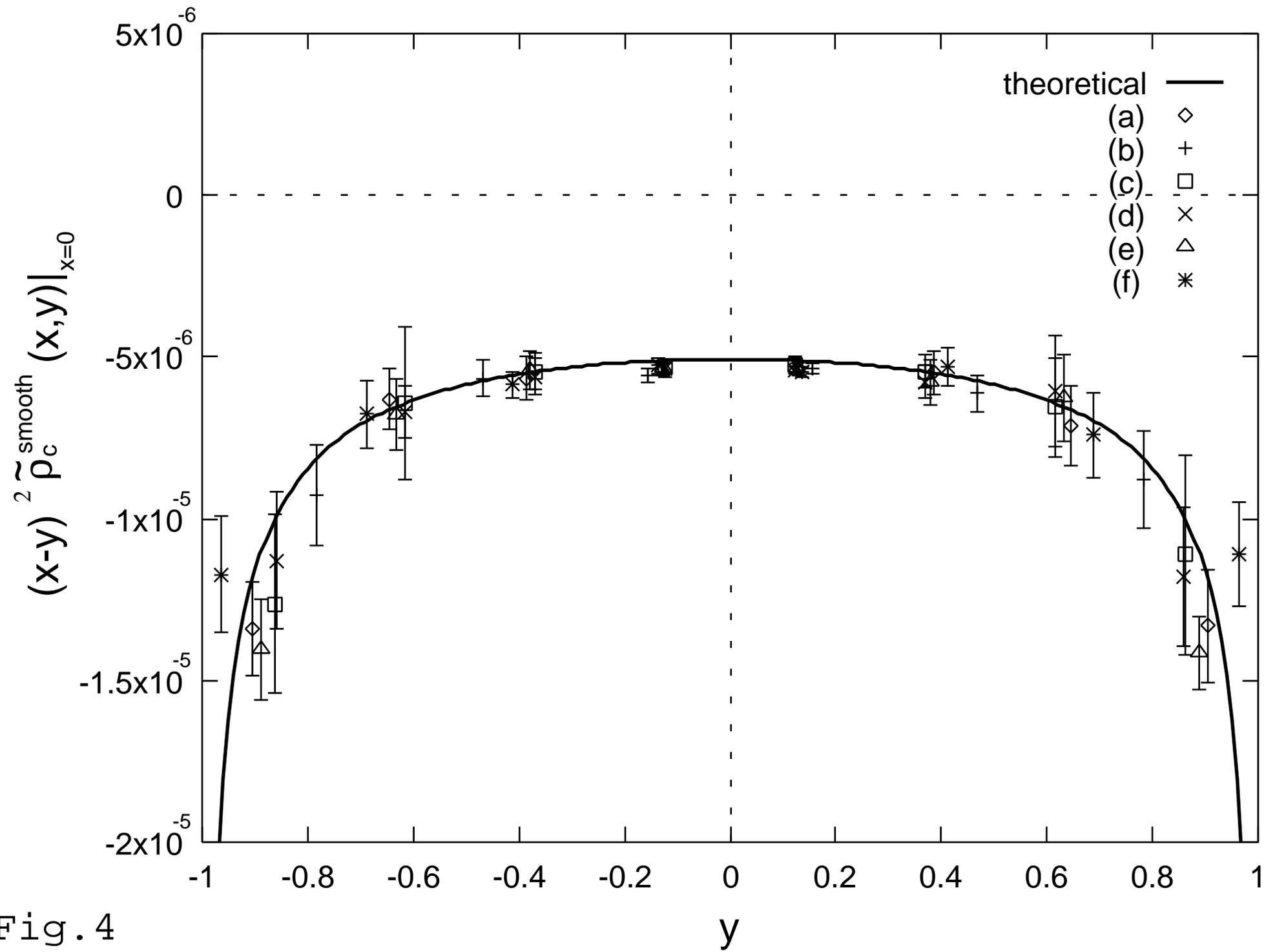
\begin{figure}
  %\epsfile{file=uc1.eps,width=8.5cm}
  \caption{The smoothed connected correlation
    $(x-y)^2\tilde{\rho}_c^{\rm \ smooth}(x,y)|_{x=0}$,
    where $x$ and $y$ are the scaling variables.
    The parameters $(v_2,~v_4,~v_6)$ are the same as in Fig.~1.
    Solid line is the theoretical result~(5).}
  \label{fig:uc1}
\end{figure}

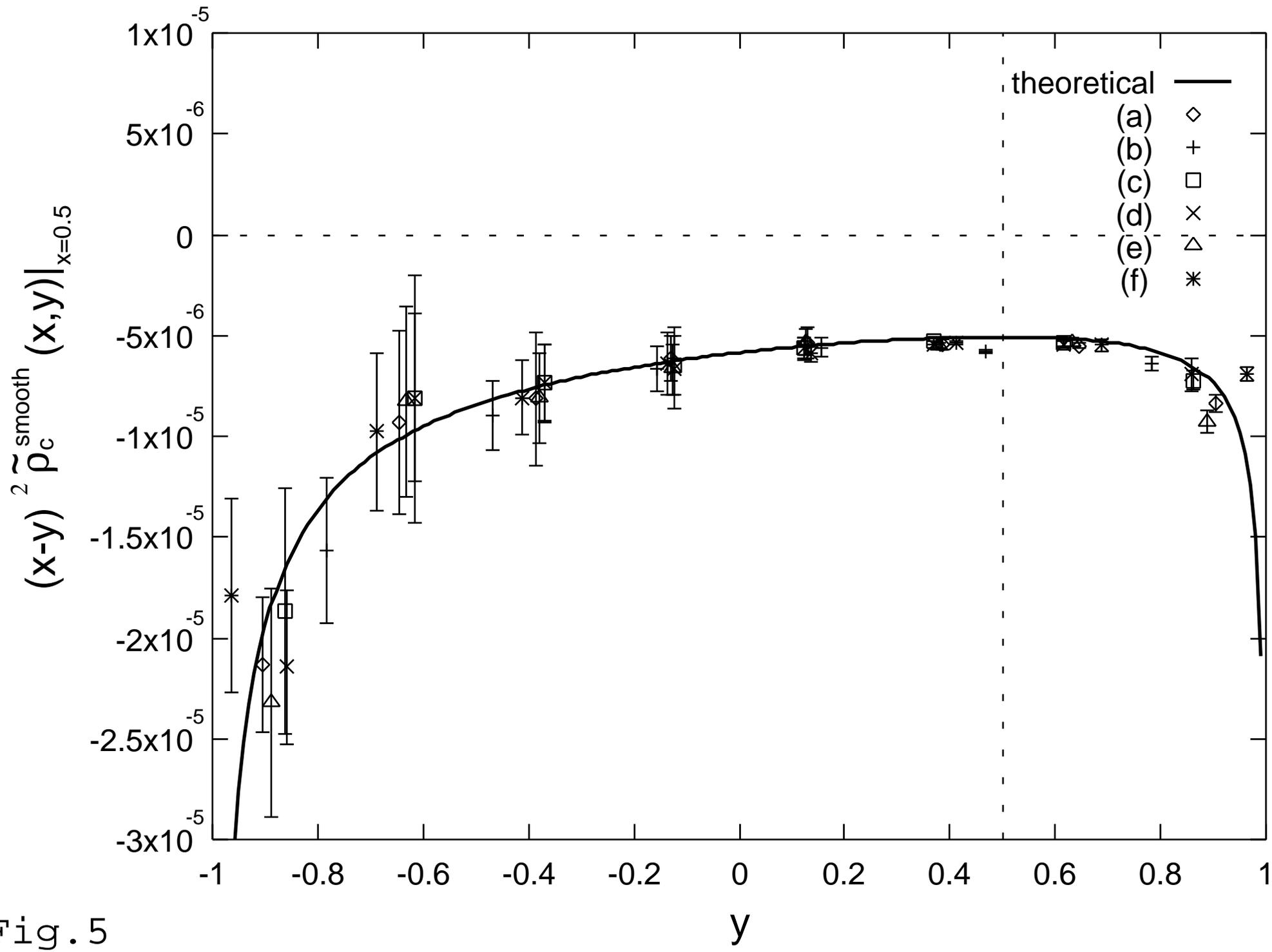
\begin{figure}
  %\epsfile{file=uc2.eps,width=8.5cm}
  \caption{The smoothed connected correlation
    $(x-y)^2\tilde{\rho}_c^{\rm \ smooth}(x,y)|_{x=0.5}$,
    where $x$ and $y$ are the scaling variables.
    The parameters $(v_2,~v_4,~v_6)$ are the same as in Fig.~1.
    Solid line is the theoretical result~(5).}
  \label{fig:uc2}
\end{figure}

First, we calculate for six sets of $v_2, v_4$ and $v_6$.
The result for the density is displayed in Fig.~\ref{fig:dens},
which shows clearly that the density is not universal.
Next, we show the smoothed correlations in Figs.~\ref{fig:c1} and \ref{fig:c2}.
Again, they are clearly not universal.
Now let us multiply the smoothed correlation $\rho_c^{\rm smooth}(\mu,\nu)$
by $a^2$ and express the result in terms of the scaling variables
$x\equiv\mu/a$ and $y\equiv\nu/a$.
The smoothed connected correlation is newly defined by the scaling variables as
\begin{equation}
  \tilde{\rho}_c^{\rm \ smooth}(x,y) = a^2 \rho_c^{\rm smooth}(\mu,\nu)
  \label{eq.rho_tild}
\end{equation}
Quite dramatically, all the results for $\tilde{\rho}_c^{\rm \ smooth}(x,y)$
fall on the same universal curve, as shown in Figs.~\ref{fig:uc1}
and \ref{fig:uc2}.

Br\'ezin and Zee\cite{pa.bz2} have also considered more general probability
distribution with even polynomials $V(\phi)$ and $W(\phi)$ as
\begin{equation}
  P(\phi) \propto \exp(-N\mathop{\rm{Tr}} V(\phi)-(\mathop{\rm{Tr}} W(\phi))^2)
  \label{la.gdist}
\end{equation}
They showed the ensemble with the distribution~(\ref{la.gdist}) can be mapped
to that of the distribution~(\ref{la.dist}) in the large $N$ limit.
Thus the smoothed connected correlation should also obey
the universal behavior~(\ref{la.univ}).

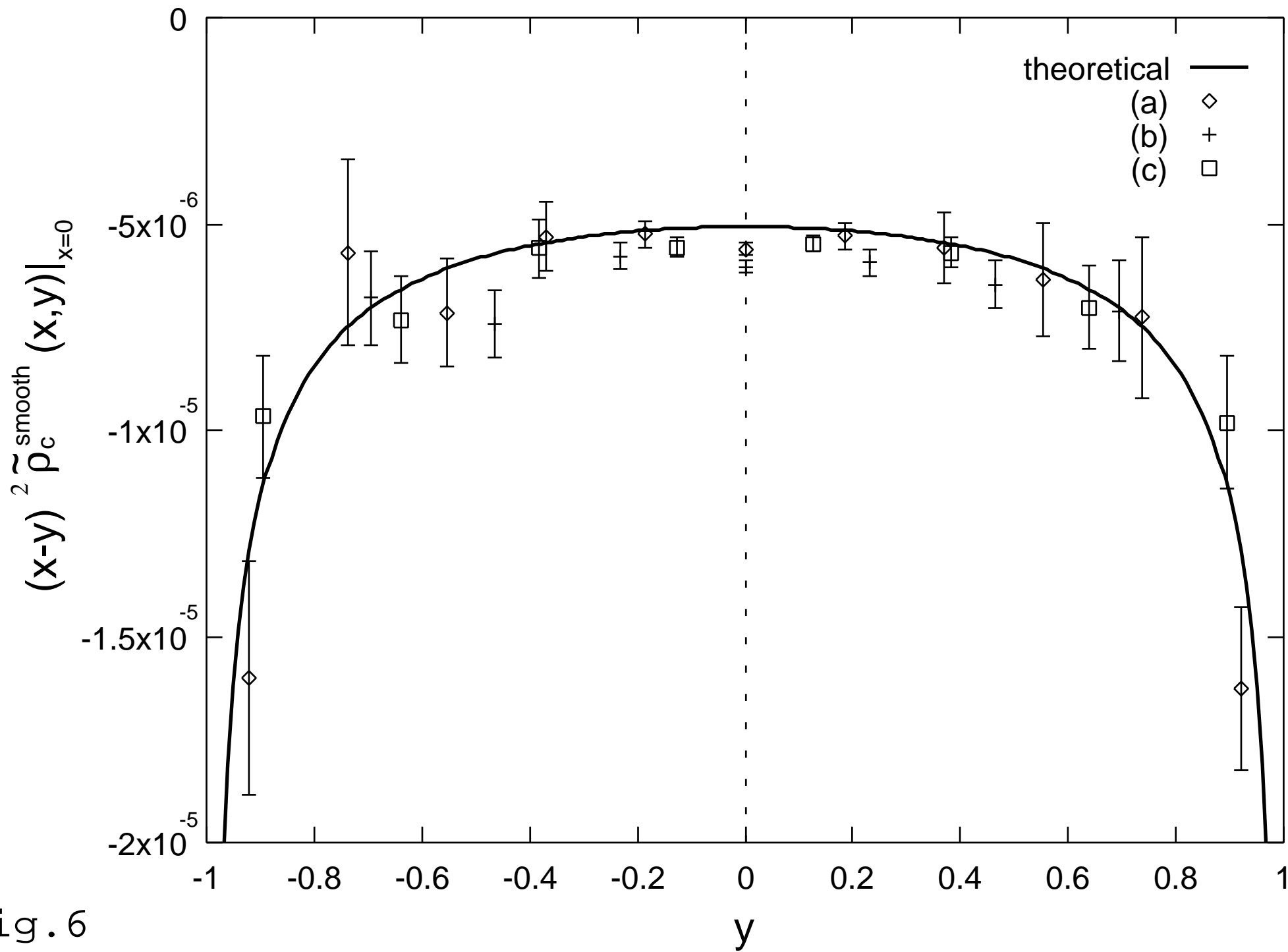
\begin{figure}
  %\epsfile{file=uc3.eps,width=8.5cm}
  \caption{The smoothed connected correlation
    $(x-y)^2\tilde{\rho}_c^{\rm \ smooth}(x,y)|_{x=0}$
    for the probability distribution~(9),
    where $x$ and $y$ are the scaling variables.
    The parameters $(v_2,~v_4,~v_6,~w_2,~w_4,~w_6)$ are
    :(a)~(1.23,~7.90,~0.290,~-0.385,~2.57,~2.80),
    (b)~(3.19,~-5.35,~5.66,~-6.28,~-9.55,~6.77)
    and (c)~(7.45,~5.67,~8.94,~3.14,~-0.0231,~8.75).
    Solid line is the theoretical result~(5).}
  \label{fig:uc3}
\end{figure}

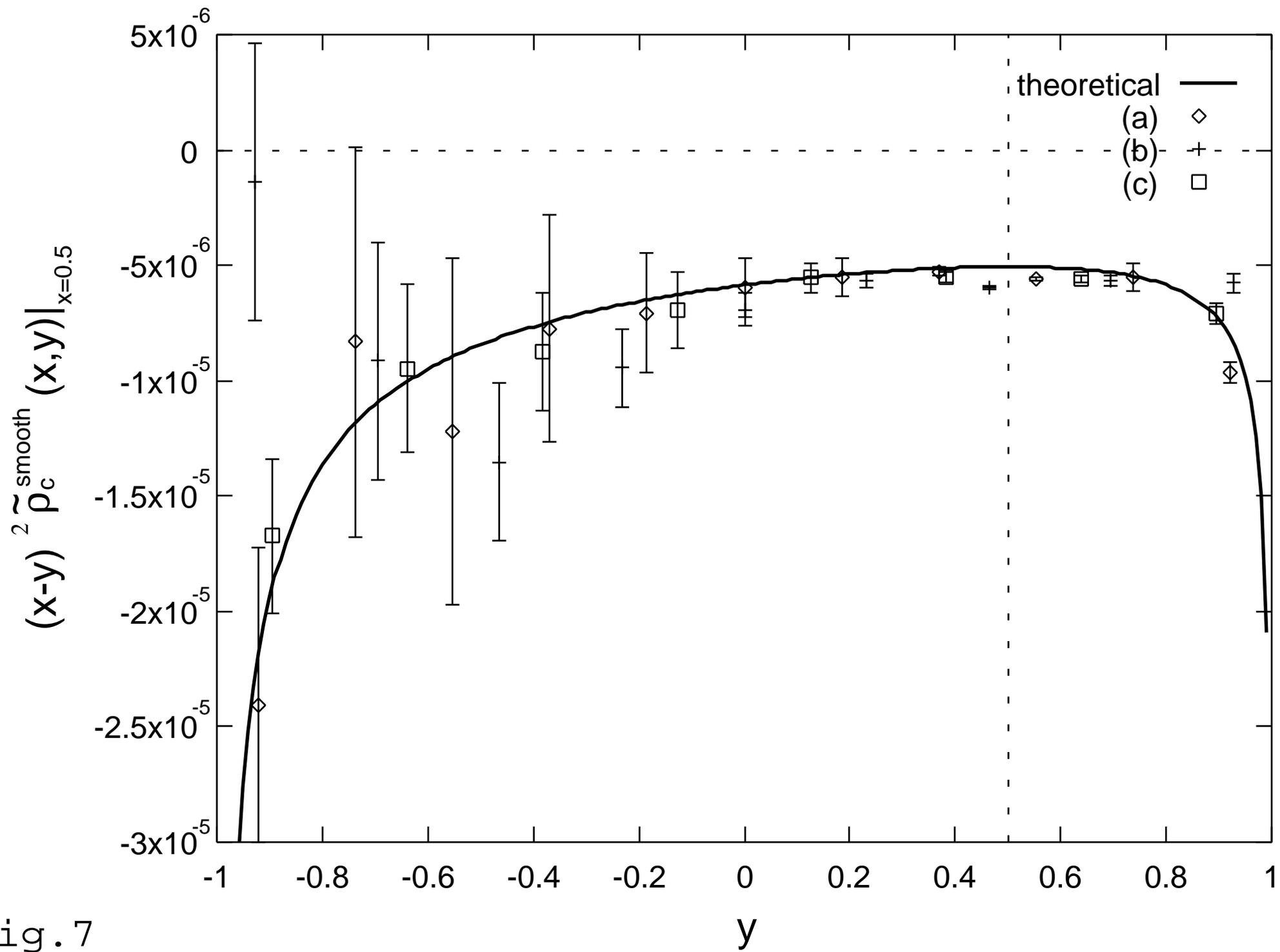
\begin{figure}
  %\epsfile{file=uc4.eps,width=8.5cm}
  \caption{The smoothed connected correlation
    $(x-y)^2\tilde{\rho}_c^{\rm \ smooth}(x,y)|_{x=0.5}$
    for probability distribution~(9),
    where $x$ and $y$ are the scaling variables.
    The parameters $(v_2,~v_4,~v_6,~w_2,~w_4,~w_6)$
    are the same as in Fig.~6.
    Solid line is the theoretical result~(5).}
  \label{fig:uc4}
\end{figure}

In this case, we also check the universal behavior numerically.
We limit the form of $V(\phi)$ and $W(\phi)$ to
\begin{eqnarray}
  V(\phi) &=& v_2 \phi^2 + v_4 \phi^4 + v_6 \phi^6 \nonumber \\
  W(\phi) &=& w_2 \phi^2 + w_4 \phi^4 + w_6 \phi^6
  \label{la.VW}
\end{eqnarray}
We choose $N$ to be 100 and take $10^6$ MC samples.
We calculate for three sets of these parameters.
The results are shown in Figs.~\ref{fig:uc3} and \ref{fig:uc4}.

We take smoothing intervals $\delta\mu$ and $\delta\nu$ to be around 0.25,
which satisfy the smoothing conditions mentioned above.
Then the dominant factor of errors is due to the MC sampling.
The errors are large near the endpoint of the spectra.
The errors grow as $|\mu-\nu|$ goes large,
because the number of the eigenvalues between $\mu$ and $\nu$ increases.
There are two regions
where the theoretical result~(\ref{la.univ}) is not applicable.
One is near the endpoint of the spectra of the eigenvalues
and the other is where two eigenvalues $\mu$ and $\nu$ are close.
Except in the two regions,
the numerical calculations for the distribution~(\ref{la.dist}) agree very
well with the theoretical correlation~(\ref{la.univ})
as shown in Figs.~\ref{fig:uc1} and \ref{fig:uc2}.
The results for the distribution~(\ref{la.gdist}) are not so good
as those for the distribution~(\ref{la.dist}) but consistent with
(\ref{la.univ})
(Figs.~\ref{fig:uc3} and \ref{fig:uc4}).

The numerical calculations show that
the smoothed connected correlation is universal
when the eigenvalues are rescaled by the endpoint
of the spectra of the eigenvalues,
although the density of the eigenvalues
and the bare correlation of the eigenvalues are not universal.
Thus we conclude that the numerical results agree well with
the universal behavior of the correlation function~(\ref{la.univ})
by Br\'ezin and Zee.
Thus the ansatz used in their derivation is well established
within the numerical errors.

Morita {\it et. al.}\cite{pa.morita} showed that the density of the eigenvalues
are separated into two parts in some parameter region.
Moreover within one part, the connected correlations of the eigenvalues
was shown to obey the universal behavior.
The numerical calculations in this parameter region agree with
the theoretical prediction to the extent that they agreed
in the previous non-separated cases.

There are also discussions that the correlations are universal for
the Wigner ensemble.\cite{pa.bz3}
The Wigner ensemble\cite{pa.wigner} is not unitary invariant
and each element of the matrix is independent and takes 1 or -1.
Our numerical calculations for the Wigner ensemble do not agree well with
the correlation function~(\ref{la.univ}).

It is a pleasure to thank Y.~Avishai for useful discussions.

\end{document}